\documentclass[12pt]{article}
\usepackage{amsmath}
\usepackage{amssymb}
\usepackage{latexsym}
\usepackage{hyperref}
\usepackage{rotating}

\def\Bra#1{\mathinner{\langle{#1}|}}
\def\Ket#1{\mathinner{|{#1}\rangle}}

\def\bra#1{\left<#1\right|}
\def\ket#1{\left|#1\right>}
{\catcode`\|=\active
  \gdef\Braket#1{\left<\mathcode`\|"8000\let|\bravert {#1}\right>}}
\def\bravert{\egroup\,\vrule\,\bgroup}
\newcommand{\alg}[1]{\mathfrak{#1}}

\newcommand{\su}{\alg{su}}
\newcommand{\Sl}{\alg{sl}}
\newcommand{\so}{\alg{so}}

\newcommand{\be}{\begin{eqnarray}}
\newcommand{\ee}{\end{eqnarray}}
\newcommand{\bea}{\begin{eqnarray}}
\newcommand{\eea}{\end{eqnarray}}
\newcommand{\ben}{\begin{equation}}
\newcommand{\een}{\end{equation}}

\newcommand{\nn}{\nonumber}

\numberwithin{equation}{section}

\begin{document}

\begin{titlepage}
\begin{flushright}
CALT-68-2512\\
hep-th/0407240
\end{flushright}
\vspace{15 mm}
\begin{center}
{\huge  $N$-Impurity superstring spectra \\
        near the pp-wave limit  }
\end{center}
\vspace{12 mm}

\begin{center}
{\large
Tristan McLoughlin and
Ian Swanson }\\
\vspace{3mm}
California Institute of Technology\\
Pasadena, CA 91125, USA
\end{center}
\vspace{5 mm}
\begin{center}
{\large Abstract}
\end{center}
\noindent

The complicated non-linear sigma model that characterizes the first finite-radius
curvature correction to the pp-wave limit of IIB superstring theory
on $AdS_5\times S^5$ has been shown to generate energy spectra that
perfectly match, to two loops in the modified 't~Hooft parameter $\lambda'$,
finite $R$-charge corrections to anomalous dimension spectra of large-$R$
${\cal N}=4$ super Yang-Mills theory in the planar limit.
This test of the AdS/CFT correspondence has been carried out for the
specific cases of two and three string excitations, which are
dual to gauge theory $R$-charge impurities.
We generalize this analysis on the string side by directly computing
string energy eigenvalues in certain protected sectors of the theory for an
arbitrary number of worldsheet excitations with arbitrary mode-number assignments.
While our results match all existing gauge theory predictions to two-loop
order in $\lambda'$, we again observe a mismatch at three loops between
string and gauge theory.  We find remarkable agreement to \emph{all}
loops in $\lambda'$, however, with the near pp-wave limit of a recently proposed
Bethe ansatz for the quantized string Hamiltonian in the $\su(2)$ sector.
Based on earlier two- and three-impurity results, we also infer the full multiplet
decomposition of the $N$-impurity superstring theory with distinct
mode excitations to two loops in $\lambda'$.

\vspace{1cm}
\begin{flushleft}
\today
\end{flushleft}
\end{titlepage}
\newpage
\section{Introduction}

The AdS/CFT correspondence is currently the subject of a new generation of
tests which endeavor to compare, near certain simplifying limits, the perturbative
anomalous dimension spectrum of planar (large-$N_c$) ${\cal N}=4$ $SU(N_c)$
super Yang-Mills (SYM) theory with the energy spectrum of
non-interacting IIB superstring theory on $AdS_5\times S^5$.
This line of investigation was sparked by the discovery that a certain large $R$-charge
limit of the gauge theory, the so-called BMN limit, corresponds
to a Penrose limit of the string background geometry in which the metric
of $AdS_5\times S^5$ is reduced to that of a pp-wave \cite{Berenstein:2002jq},
a background in which the string theory is free in lightcone gauge
\cite{Metsaev:2001bj,Metsaev:2002re}.  This matching becomes much more
elaborate when higher-order perturbative corrections in the 't~Hooft
coupling $\lambda = g_{\rm YM}^2 N_c$ are included in the gauge theory
\cite{Beisert:2003tq,Beisert:2003jb,Beisert:2003ys,Serban:2004jf,Beisert:2004hm},
or when spacetime curvature corrections away from the pp-wave limit are
admitted in the string theory
\cite{Parnachev:2002kk,Callan:2003xr,Callan:2004uv,Callan:2004ev,Swanson:2004mk}.
Since the Penrose limit of the string theory is realized by boosting string
states along an equatorial geodesic in the $S^5$ subspace, the latter corrections to the
pp-wave geometry emerge in inverse powers of the $S^5$ string angular momentum $J$.

A number of important discoveries have advanced the gauge theory
side of these studies.  It has been realized, for example, that
certain sectors of the ${\cal N}=4$ dilatation generator can be
mapped to one-dimensional integrable spin-chain Hamiltonians
\cite{Minahan:2002ve}.  These sectors are typically labelled by
the subalgebra of the full $\alg{psu}(2,2|4)$ superconformal
algebra under which they are invariant, and they can be modelled
by spin-chain systems which are invariant under the same symmetry.
The problem of computing operator anomalous dimensions in a given
sector of the gauge theory can therefore be replaced with that of
finding the energy eigenvalue spectrum of a corresponding spin
chain.  The integrability of these spin-chain systems implies that
their energy spectra can be extracted via techniques such as the
Bethe ansatz.  This powerful tool was first applied in this
context at one-loop order ($O(\lambda)$) in an $\so(6)$ sector by
Minahan and Zarembo \cite{Minahan:2002ve}, and generalized to the
full superconformal symmetry algebra in \cite{Beisert:2003yb}.
(The Bethe ansatz approach to integrable systems is described in a
more general setting in \cite{Faddeev:1996iy}, for example.) The
Bethe equations are usually exactly soluble for spin chains with
two impurities (the spin-chain impurity number refers to the
number of $R$-charge defects in the corresponding SYM operator,
and is equivalent to the number of worldsheet excitations in the
dual string theory).  For higher impurity number, the one-loop
Bethe equations can be solved perturbatively near the limit of
infinite chain length, a limit which corresponds to the large-$J$,
pp-wave limit of the string theory. Furthermore, since
integrability has been shown to persist in the gauge theory to at
least three-loop order (in an $\su(2)$ sector), ``long-range''
Bethe equations have emerged encoding the higher-loop dynamics of
the theory \cite{Serban:2004jf,Beisert:2004hm}.  By incorporating
non-nearest-neighbor interactions of pseudoparticle excitations on
the spin lattice that are typical of higher-loop SYM spin-chain
Hamiltonians, such equations are in fact meant to describe the
gauge theory physics to \emph{all orders} in $\lambda$ (the
Inozemtsev chain of \cite{Serban:2004jf} failed to exhibit proper
BMN scaling at four loops, but this shortcoming was circumvented
by the modified ansatz of \cite{Beisert:2004hm}). Predictions from
the long-range $\su(2)$ Bethe equations have recently been tested
against a separate virial technique for operators with three
$R$-charge impurities in \cite{Virial}, where agreement was
obtained to three-loop order. (The $\su(2)$ spin-chain Hamiltonian
has not been fixed definitively
beyond this order.)

On the string theory side the first $1/J$ correction to the free
pp-wave spectrum was computed in
\cite{Parnachev:2002kk,Callan:2003xr,Callan:2004uv}. The string
energies in this setting correspond in the gauge theory to the
difference between operator scaling dimensions and $R$-charge
($\Delta \equiv D-R$), and states are arranged into superconformal
multiplets according to the $\alg{psu}(2,2|4)$ symmetry of the
theory. The fully supersymmetric two-excitation (or two-impurity)
system, for example, is characterized by a 256-dimensional
supermultiplet of states built on a scalar primary. The complete
spectrum of this system was successfully matched to corresponding
SYM operator dimensions in \cite{Callan:2003xr,Callan:2004uv} to
two loops in the modified 't~Hooft coupling $\lambda' =
\lambda/J^2$. A three-loop mismatch between the gauge and string
theory results discovered therein comprises a long-standing and
open problem in these studies, one which has appeared in several
different contexts (see,
eg.,~\cite{Serban:2004jf,Beisert:2004hm,Beisert:2003ea}). This was
extended to the three-impurity, 4,096-dimensional supermultiplet
of string states in \cite{Callan:2004ev}, where precise agreement
with the corresponding gauge theory was again found to two-loop
order, and a general disagreement reappeared at three loops. In
the latter study, three-impurity string predictions were compared
with corresponding gauge theory results derived both from the
virial technique described in \cite{Virial} and the long-range
Bethe ansatz of \cite{Beisert:2003yb} (which overlaps at one loop
with the original $\so(6)$ system studied in
\cite{Minahan:2002ve}).

In the present study we generalize the string side of these investigations
by computing, directly from the Hamiltonian, various $N$-impurity spectra
of IIB superstring theory at $O(J^{-1})$ in the large-$J$ curvature expansion near
the pp-wave limit of $AdS_5\times S^5$.
We focus on the bosonic $\su(2)$ and $\Sl(2)$
sectors which are characterized by $N$ symmetric-traceless
bosonic string excitations in the $S^5$ and $AdS_5$ subspaces, respectively.
These sectors are known to decouple from the theory to all orders in $\lambda'$,
and are thus referred to as ``closed'' sectors of the theory.
Based on calculations in the $\su(2)$ and $\Sl(2)$ sectors,
we also formulate a conjecture for the
$N$-impurity spectrum of states in a protected $\su(1|1)$ sector composed of $N$ fermionic
excitations symmetrized in their $SO(4)\times SO(4)$ spinor indices.
We then describe the complete supermultiplet decomposition of the $N$-impurity
spectrum to two loops in $\lambda'$ using a simple generalization of the two- and
three-impurity cases.

We note that a new Bethe ansatz for the string theory was recently proposed
by Arutyunov, Frolov and Staudacher \cite{Arutyunov:2004vx} which is meant
to diagonalize the fully quantized string sigma model in the $\su(2)$ sector
to all orders in $1/J$ and $\lambda'$.
This ansatz was shown in \cite{Arutyunov:2004vx} to
reproduce the two- and three-impurity
spectra of quantized string states near the pp-wave limit detailed in
\cite{Callan:2004uv,Callan:2004ev}.  The methods developed here allow us to
check their formulas directly against the string theory for any impurity number
at $O(J^{-1})$,
and we find that our general $\su(2)$ string eigenvalues
agree to all orders in $\lambda'$
with their $\su(2)$ string Bethe ansatz!

Section 2 is dedicated to a brief review of the string system near the
pp-wave limit.  We compute the $N$-impurity energy spectra of the $\su(2)$, $\Sl(2)$
and $\su(1|1)$ closed sectors of this system in section 3, and generalize the complete
$N$-impurity supermultiplet structure of the theory to two-loop order in
$\lambda'$ in section 4.  We conclude in section 5 with a discussion of
future problems.

\section{Notation and string quantization in $AdS_5\times S^5$}
For convenience we will present the first curvature correction to
the bosonic sector of the string Hamiltonian near the pp-wave limit
of $AdS_5\times S^5$, briefly review its derivation and introduce
some standard notation.
The spacetime metric of $AdS_5\times S^5$ can be written as
\be
ds^2 = R^2\left[
    -\left(\frac{1+\frac{1}{4}z^2}{1-\frac{1}{4}z^2}\right)^2 dt^2
    +\left(\frac{1-\frac{1}{4}y^2}{1+\frac{1}{4}y^2}\right)^2 d\phi^2
    +\frac{dz_k dz_k}{(1-\frac{1}{4}z^2)^2}
    +\frac{dy_{k'} dy_{k'}}{(1+\frac{1}{4}y^2)^2}
    \right]\ ,
\ee
where $R$ is the spacetime radius and the coordinates $z_k$ ($k=1,\ldots,4$)
and $y_{k'}$ ($k'=5,\ldots,8$) parameterize two transverse $SO(4)$ spaces
which descend from the $AdS_5$ and $S^5$ subspaces, respectively.  Lightcone
coordinates are introduced by the reparameterization
\be
t=x^+ \qquad \phi = x^+ + x^-/R^2\ ,
\ee
with corresponding momenta
\be
-p_+ = \Delta-J \qquad -p_-=i\partial_{x^-} = \frac{i}{R^2}\partial_\phi
    = -\frac{J}{R^2}\ .
\ee
The $S^5$ angular momentum $J$ is related to the scale factor $R$ by
$p_-R^2 = J$.  To reach the pp-wave limit, the eight transverse coordinates
$z_k$ and $y_{k'}$ are rescaled according to
\be
z_k \to z_k/R \qquad y_{k'} \to y_{k'}/R\ ,
\ee
and $p_-$ is held fixed while $R$
and $J$ are taken to be infinite.  Under the AdS/CFT parameter
duality, $p_-$ is given by
\be
p_- = \frac{1}{\sqrt{\lambda'}} = \frac{J}{\sqrt{g_{\rm YM}^2 N_c}}\ ,
\ee
where $\lambda'$ is the so-called modified 't~Hooft coupling.
Keeping the first $1/R^2$ (equivalently $1/J$)
correction to this limit, the metric becomes
\be
ds^2 & = & 2\,dx^+ dx^- - (x^A)^2(dx^+)^2+(dx^A)^2
\nn\\
&&  +\frac{1}{R^2}\left[
    -2y^2 dx^+ dx^-+\frac{1}{2}(y^4-z^4)(dx^+)^2+(dx^-)^2
    +\frac{1}{2}z^2dz^2-\frac{1}{2}y^2dy^2\right]
\nn\\
&&  +O(R^{-4})\ ,
\ee
where the coordinates $x^A$ ($A=1,\ldots,8$)
span the transverse $SO(8)$ subspace.

The complete IIB superstring theory in this background can be formulated
in terms of the Green-Schwarz action built from
supersymmetric Cartan one-forms and superconnections on the
coset space associated with $AdS_5\times S^5$
\cite{Kallosh:1998zx,Kallosh:1998ji,Metsaev:1998it,Metsaev:1999gz,Metsaev:2000yf}.
The complete Green-Schwarz superstring Lagrangian, which is dependent on
the scale radius $R$, can then be expanded in large $R$
to extract the pp-wave limit of the theory plus higher-order corrections in $R^{-2}$.
At leading order in this expansion the Hamiltonian $H_{\rm pp}$
consists of a free theory of eight massive bosons and fermions:
\begin{eqnarray}
{H}_{\rm pp} & = &
    \frac{p_-}{2}(x^A)^2 + \frac{1}{2p_-}\left[(p_A)^2 + ({x'}^A)^2\right]
    + {i}\rho\Pi\psi + \frac{i}{2}\psi\psi' - \frac{i}{2p_-^2} \rho \rho'\ .
\end{eqnarray}
The fields $p_A$ are bosonic momenta conjugate to $x^A$, $\psi$
denote fermionic fields with conjugate variables $\rho$, and
the shorthand notation ${x'}^A$ denotes the worldsheet derivative
$\partial_\sigma x^A$.  The fermion fields $\psi_\alpha$ are eight-component
complex spinors constructed from two $SO(9,1)$ Majorana-Weyl spinors of
equal chirality, and the matrix $\Pi$ is defined in terms of the eight-dimensional
$SO(8)$ gamma matrices $\gamma^a,\ \bar\gamma^a$ as
$\Pi \equiv \gamma^1\bar\gamma^2\gamma^3\bar\gamma^4$.
(For further details the reader is referred to \cite{Callan:2003xr,Callan:2004uv}.)

The first curvature correction to the background metric gives rise to an
interaction Hamiltonian denoted by $H_{\rm int}$, which, in turn, is broken
into bosonic $(H_{\rm BB})$, fermionic $(H_{\rm FF})$ and bose-fermi mixed
$(H_{\rm BF})$ sectors:
\begin{eqnarray}
\label{GSHamFinal}
{H}={H}_{\rm pp}+{H}_{\rm int}  \qquad
{H}_{\rm int}={H}_{\rm BB}+{H}_{\rm FF}+{H}_{\rm BF}~.
\end{eqnarray}
The complete perturbation $H_{\rm int}$ was computed explicitly in terms of the
constituent fields described above in \cite{Callan:2003xr,Callan:2004uv}.
Since we will deal mostly with the bosonic sector
of the theory in the present study, we state $H_{\rm BB}$ explicitly but refer
the reader to \cite{Callan:2003xr,Callan:2004uv} for the detailed form of
the remaining sectors $H_{\rm BF}$ and $H_{\rm FF}$:
\be
\label{Hpurbos}
{H}_{\rm BB} & = & \frac{1}{R^2}\biggl\{
    \frac{1}{4p_-}\left[ -y^2\left( p_z^2 + {z'}^2 + 2{y'}^2\right)
    + z^2\left( p_{y}^2 + {y'}^2 + 2{z'}^2 \right)\right]
    + \frac{p_-}{8}\left[ (x^A)^2 \right]^2
\nn\\
& &     - \frac{1}{8p_-^3}\left\{  \left[ (p_A)^2\right]^2 + 2(p_A)^2({x'}^A)^2
    + \left[ ({x'}^A)^2\right]^2 \right\}
     + \frac{1}{2p_-^3}\left({x'}^A p_A\right)^2
    \biggr\}\ .
\ee
The $SO(8)$ indices $(A,B,C,\ldots=1,\ldots,8)$ are generally
split into  $AdS_5$ and $S^5$ subspaces using the lower-case Latin notation
$a,b,c,\ldots = 1,\ldots,4$ (for $SO(4)_{AdS}$) and
$a',b',c',\ldots = 5,\ldots,8$ (for $SO(4)_{S^5}$).
The Greek indices $\alpha,\beta,\gamma,\ldots = 1,\ldots,8$
are used to label the eight components of the fermionic fields $\psi,\ \rho$.
We point out that the interaction Hamiltonian is strictly quartic in fields,
a fact which will be important in the subsequent analysis.

The vacuum state carries the $S^5$ string angular momentum $J$ and
is labelled by $\ket{J}$; the complete Fock space of string states
is generated by acting on $\ket{J}$ with any number of the
creation operators $a_n^{A\dag}$ (bosonic) and $b_n^{\alpha\dag}$
(fermionic), where the lower indices $n,m,l,\ldots$ denote mode
numbers. The excitation number of string states (defined by the
number of creation oscillators acting on the ground state) will
also be referred to as the impurity number, and string states with
a total of $N_B+N_F = N$ impurities will contain $N_B$ bosonic and
$N_F$ fermionic impurities: \be \Ket{N_B,N_F;J} \equiv
\underset{N_B}{\underbrace{a_{n_1}^{A_1\dag}a_{n_2}^{A_2\dag}
    \ldots a_{n_{N_B}}^{A_{N_B}\dag}}}
    \underset{N_F}{\underbrace{b_{n_1}^{\alpha_{1}\dag}b_{n_2}^{\alpha_2\dag}
    \ldots b_{n_{N_F}}^{\alpha_{N_F}\dag} }}\ket{J}\ .
\ee
States constructed in this manner fall into two disjoint subsectors populated by
spacetime bosons ($N_F$ even) and spacetime fermions ($N_F$ odd).  In this notation
the pure-boson states $\ket{N_B,0;J}$ are mixed only by $H_{\rm BB}$ and
the pure-fermion states $\ket{0,N_F;J}$ are acted on by $H_{\rm FF}$.  The more
general spacetime-boson states $\ket{N_B,{\rm even};J}$ are acted on by the
complete interaction Hamiltonian $H_{\rm int}$, as are the spacetime-fermion states
$\ket{N_B,{\rm odd};J}$.  There is of course no mixing between spacetime bosons and
fermions; this block-diagonalization is given schematically in table~\ref{block}.
\begin{table}[ht!]
\begin{eqnarray}
\begin{array}{|c|cccc|}
\hline
 H_{\rm int} & \Ket{N_B,0;J} &
        \Ket{N_B,{\rm even};J} &
        \Ket{N_B,{\rm odd};J} &
        \Ket{0,{\rm odd};J}
\\   \hline
\Bra{N_B,0;J}   & H_{\rm BB}    & H_{\rm BF}                &   &   \\
\Bra{N_B,{\rm even};J} & H_{\rm BF} & H_{\rm BB}+H_{\rm BF}+H_{\rm FF}  &   &   \\
\Bra{N_B,{\rm odd};J} &  &   & H_{\rm BB}+H_{\rm BF}+H_{\rm FF} & H_{\rm BF} \\
\Bra{0,{\rm odd};J} &  &  & H_{\rm BF} & H_{\rm FF}  \\
\hline
\end{array} \nonumber
\end{eqnarray}
\caption{Interaction Hamiltonian on $N$-impurity string states $(N_B+N_F=N)$}
\label{block}
\end{table}

The full interaction Hamiltonian can be further block-diagonalized
by projecting onto certain protected sectors of string states, and
we will focus in this study on three such sectors.  Two of these
sectors are spanned by purely bosonic states $\ket{N_B,0;J}$
projected onto symmetric-traceless irreps in either the
$SO(4)_{AdS}$ or $SO(4)_{S^5}$ subspaces.  Another sector which is
known to decouple at all orders in $\lambda'$ is comprised of
purely fermionic states $\ket{0,N_F;J}$ projected onto either of
two subspaces of $SO(4)\times SO(4)$ labelled, in an
$SU(2)^2\times SU(2)^2$ notation, by $({\bf 2,1;2,1})$ and $({\bf
1,2;1,2})$, and symmetrized in spinor indices. Each of these
sectors can also be labelled by the subalgebra of the full
superconformal algebra that corresponds to the symmetry under
which they are invariant.  The bosonic $SO(4)_{AdS}$ and
$SO(4)_{S^5}$ sectors are labelled by $\Sl(2)$ and $\su(2)$
subalgebras, respectively, while the two fermionic sectors fall
into $\su(1|1)$ subsectors of the closed $\su(2|3)$ system studied
in \cite{Beisert:2003ys,Beisert:2003yb,Beisert:2003jj}.

\section{$N$-Impurity string energy spectra}
In the large-$J$ expansion about the free pp-wave theory,
we will isolate $O(J^{-1})$ corrections to the energy eigenvalues of
$N$-impurity string states according to
\be
E(\{q_j\},N,J) = \sum_{j=1}^N\sqrt{1+q_j^2\lambda'} + \delta E(\{q_j\},N,J) + O(J^{-2})\ .
\ee
The spectrum is generically dependent upon $\lambda'$, $J$ and
the mode numbers $\{n_j\},\{q_j\},\ldots,$ where $j$ is understood
to label either the complete set of impurities ($j=1,\ldots,N$) or
some subset thereof (eg.~$j=1,\ldots,N_F$).
The leading order term in this expansion is the $N$-impurity free energy of states
on the pp-wave geometry, and $\delta E(\{q_j\},N,J)$ always
enters at $O(J^{-1})$.
When it becomes necessary, we will also expand the $O(1/J)$ energy shift
in the small-$\lambda'$ loop expansion:
\be
\delta E(\{q_j\},N,J) = \sum_{i=1}^\infty \delta E^{(i)}(\{q_j\},N,J) ({\lambda'})^{i}\ .
\ee
Finding the explicit form of $\delta E(\{q_j\},N,J)$
for $N$-impurity string states in certain interesting sectors of the theory will be
our primary goal.  As a side result, however, we will see that the spectrum
of \emph{all} states in the theory will be determined to two-loop order
in $\lambda'$ by the specific eigenvalues we intend to compute.

We begin by noting that the canonical commutation relations
of the bosonic fields $x^A$ and $p_A$ allow us to expand
$H_{\rm BB}$ in bosonic creation and annihilation operators using
\be
x^A(\sigma,\tau)  &=&  \sum_{n=-\infty}^\infty x_n^A(\tau ) e^{-i k_n\sigma}  \nn\\
x_n^A(\tau) &=& \frac{i}{\sqrt{2\omega_n}}\left(a_n^A e^{-i\omega_n\tau}
        -a_{-n}^{A\dag}e^{i\omega_n\tau}\right)\ ,
\ee
where $k_n = n$ are integer-valued, $\omega_n=\sqrt{p_-^2 + n^2}$ and
the operators $a_n^A$ and $a_n^{A\dag}$ obey the usual relation
$\left[a_m^A,a_n^{B\dag}\right] = \delta_{mn}\delta^{AB}$.
Since we are only interested in computing diagonal matrix elements of $H_{\rm BB}$
between physical string states with equal numbers of excitations, we can
restrict the oscillator expansion to contain only equal numbers of
creation and annihilation operators (all other combinations automatically
annihilate between equal-impurity string states).
Explicitly, we obtain the following expansion:
\begin{eqnarray}
\label{Hcorrected}
{H}_{\rm BB} & = &
    -\frac{1}{32 p_- R^2}\sum \frac{\delta(n+m+l+p)}{\xi}
    \times
\nn\\
& & \biggl\{
    2 \biggl[ \xi^2
    - (1 - k_l k_p k_n k_m )
     +  \omega_n \omega_m k_l k_p
      +  \omega_l \omega_p k_n k_m
    + 2 \omega_n \omega_l k_m k_p
\nn\\
& &\kern-10pt
     + 2 \omega_m \omega_p k_n k_l
    \biggr]
    a_{-n}^{\dagger A}a_{-m}^{\dagger A}a_l^B a_p^B
   +4 \biggl[ \xi^2
    - (1 - k_l k_p k_n k_m )
     - 2 \omega_n \omega_m k_l k_p
     +  \omega_l \omega_m k_n k_p
\nn\\
& &   -  \omega_n \omega_l k_m k_p
    -  \omega_m \omega_p k_n k_l
    + \omega_n \omega_p k_m k_l \biggr]
    a_{-n}^{\dagger A}a_{-l}^{\dagger B}a_m^A a_p^B
     + 2  \biggl[8 k_l k_p
    a_{-n}^{\dagger i}a_{-l}^{\dagger j}a_m^i a_p^j
\nn\\
& &     + 2 (k_l k_p +k_n k_m)
    a_{-n}^{\dagger i}a_{-m}^{\dagger i}a_l^j a_p^j
    +(\omega_l \omega_p+ k_l k_p -\omega_n
    \omega_m- k_n k_m)a_{-n}^{\dagger i}a_{-m}^{\dagger i}a_l^{j'} a_p^{j'}
\nn\\
& &     -4 ( \omega_l \omega_p- k_l k_p)
    a_{-n}^{\dagger i}a_{-l}^{\dagger j'}a_m^i a_p^{j'}
    -(i,j \rightleftharpoons i',j')
    \biggr]\biggr\}~,
\end{eqnarray}
where $\xi \equiv \sqrt{\omega_n\omega_m\omega_l\omega_p}$.

\subsection{The $SO(4)_{S^5}$ ($\su(2)$) sector}
We begin in the $\su(2)$ sector spanned by symmetric-traceless
pure-boson states excited in the $S^5$ subspace.
Because we are restricting our attention to $SO(4)_{S^5}$ states
symmetric in their vector indices, we form the following oscillators:
\be
a_n = \frac{1}{\sqrt{2}}\left( a^5_n + i a^6_n \right) \qquad
\bar a_n = \frac{1}{\sqrt{2}}\left( a^5_n - i a^6_n \right)\ .
\label{oscdef}
\ee
By taking matrix elements of the form
\be
\bra{J} a_{n_1}a_{n_2}\ldots a_{n_{N_B}} ( H_{\rm BB} )
    a_{n_1}^\dag a_{n_2}^\dag \ldots a_{n_{N_B}}^\dag \ket{J}\ ,
\label{ME}
\ee
we can therefore select out excitations in the $(5,6)$-plane of the $S^5$ subspace and
make the symmetric-traceless projection manifest.  (More generally we
can project onto any $(n,m)$-plane, as long as $n\neq m$
and both are chosen to lie in the $S^5$ subspace.)

There are two basic oscillator structures of $H_{\rm BB}$ in eqn.~(\ref{Hcorrected}):
one in which the creation (annihilation) operators are contracted
in their $SO(4)\times SO(4)$ indices
\be
a_{-n}^{\dag A} a_{-m}^{\dag A} a_{l}^{B}a_{p}^{B}\ , \nn
\ee
and one where pairs of creation and annihilation operators are contracted
\be
a_{-n}^{\dag A} a_{-l}^{\dag B} a_{m}^{A}a_{p}^{B}\ . \nn
\ee
In terms of the $a_n$ and $\bar a_n$ fields of eqn.~(\ref{oscdef}),
the former structure contains
\be
a_{-n}^{\dag A} a_{-m}^{\dag A} a_{l}^{B}a_{p}^{B}\Bigr|_{(5,6)} =
    \bigl( a_{-n}^\dag\,\bar a_{-m}^\dag + \bar a_{-n}^\dag\, a_{-m}^\dag \bigr)
    \bigl( a_l\,\bar a_p + \bar a_l\, a_p\bigr)\ ,
\ee
which cannot contribute to $\su(2)$ matrix elements of the form appearing in
(\ref{ME}).  The latter structure, however, contains
\be
a_{-n}^{\dag A} a_{-l}^{\dag B} a_{m}^{A}a_{p}^{B}\Bigr|_{(5,6)} =
    \bar a_{-n}^\dag\, \bar a_{-l}^\dag\, \bar a_m\, \bar a_p
    + a_{-n}^\dag\, a_{-l}^\dag\, a_m\, a_p\ ,
\label{osc2}
\ee
which will contribute to the $\su(2)$ energy spectrum.

The string states appearing in the matrix element of
eqn.~(\ref{ME}) have been written in the generic form \be
a_{n_1}^\dag a_{n_2}^\dag \ldots a_{n_{N_B}}^\dag \ket{J}\ , \nn
\ee and, as usual, they are subject to the level-matching
condition \be \sum_{j=1}^{N_B} n_j = 0\ . \label{LM} \ee The
complete set of mode indices $\{n_1,n_2,\ldots,n_{N_B}\}$ can
contain one or more subsets of indices that are equal, while still
satisfying eqn.~(\ref{LM}); this scenario complicates the
calculation of energy eigenvalues to some extent. We will
eventually compute the eigenvalues of interest for completely
general string states, but for purposes of illustration and to
introduce our strategy we will start with the simplest case in
which no two mode numbers are equal $(n_1\neq n_2 \neq \ldots\neq
n_{N_B})$. To organize the presentation, we will generally use
mode numbers labelled by $\{n_j\}$ to denote those which are
inequivalent from each other, while $\{q_j\}$ will be allowed to
overlap.  Between states with completely distinct mode indices,
the oscillator structure in eqn.~(\ref{osc2}) exhibits the
following matrix element: \be &&\bra{J}a_{n_1}a_{n_2}\ldots
a_{N_B} ( a_{-n}^\dag a_{-l}^\dag a_m a_p )
    a_{n_1}^\dag a_{n_2}^\dag \ldots a_{N_B}^\dag \ket{J}
\nn\\
&&\kern+60pt
    = \frac{1}{2}
    \sum_{j,k=1 \atop j\neq k}^{N_B}
    \Bigl(
    \delta_{n_j+n}\,\delta_{n_k+l}\,\delta_{n_j-m}\,\delta_{n_k-p}
    +\delta_{n_j+n}\,\delta_{n_k+l}\,\delta_{n_k-m}\,\delta_{n_j-p}
\nn\\
&&\kern+110pt
    +\delta_{n_j+l}\,\delta_{n_k+n}\,\delta_{n_j-m}\,\delta_{n_k-p}
    +\delta_{n_j+l}\,\delta_{n_k+n}\,\delta_{n_k-m}\,\delta_{n_j-p}
    \Bigr)\ .
\label{delME1}
\ee
With this in hand, it is a straightforward exercise to compute the
energy eigenvalue of the $SO(4)_{S^5}$ bosonic interaction
Hamiltonian in the $N_B$-impurity symmetric-traceless irrep
(with unequal mode indices):  we simply attach the $H_{\rm BB}$ coefficient
of the oscillator structure $ a_{-n}^\dag a_{-l}^\dag a_m a_p $
to the right-hand side of eqn.~(\ref{delME1}) and carry out the
summation over mode numbers.  The result is remarkably compact:
\be
\delta E_{{S^5}}(\{n_i\},N_B,J) = -\frac{1}{J}\sum_{j,k=1\atop j\neq k}^{N_B}
    \frac{1}{2\,\omega_{n_j}\omega_{n_k}}\left[
    n_k^2 + n_j^2\left(1+n_k^2\lambda'\right)
    + n_j n_k\left(1-\omega_{n_j}\omega_{n_k}\lambda'\right)\right]\ .
\label{su2GEN}
\ee

This $\su(2)$ formula can be checked against
previously obtained string theory results in the two- and three-impurity regimes.
Namely, the two-impurity eigenvalue computed in \cite{Callan:2003xr,Callan:2004uv}
takes the form (which is exact in $\lambda'$)
\be
\delta E_{{S^5}}(n_1,n_2,J) = -\frac{2\,n_1^2\lambda'}{J}\ ,
\label{2impsu2}
\ee
where we have set $n_2= -n_1$ using eqn.~(\ref{LM}).
This eigenvalue matches the general formula in eqn.~(\ref{su2GEN})
restricted to two impurities.  The $\su(2)$ eigenvalue for three impurities
with unequal mode indices $(n_1\neq n_2\neq n_3)$ was calculated in
\cite{Callan:2004ev} and found to be
\be
\delta E_{{S^5}}(n_1,n_2,n_3,J) &=&
    -\frac{1}{J\omega_{n_1}\omega_{n_2}\omega_{n_3}}\biggl\{
    \left[n_1n_2+n_2^2+n_1^2(1+n_2^2\lambda')\right]\omega_{n_3}
\nn\\
&&\kern-20pt
    + \left[n_1n_3+n_3^2+n_1^2(1+n_3^2\lambda')\right]\omega_{n_2}
    + \left[n_2n_3+n_3^2+n_2^2(1+n_3^2\lambda')\right]\omega_{n_1}
\nn\\
&&\kern-20pt
    -\left[n_2n_3+n_1(n_2+n_3)\right]\lambda'\omega_{n_1}\omega_{n_2}\omega_{n_3}
    \biggr\}\ .
\label{3impsu2}
\ee
It is also easy to check that eqn.~(\ref{su2GEN})
reproduces this formula exactly for $N_B = 3$.

Since eqn.~(\ref{su2GEN}) matches all previously computed
results from the string theory in this sector,
it must therefore agree with corresponding
$\su(2)$ gauge theory predictions only to two-loop order in $\lambda$.
We note, however, that eqn.~(\ref{su2GEN}) is \emph{identical}
to the $N$-impurity $O(J^{-1})$ energy shift (with unequal mode numbers)
obtained from the $\su(2)$ string Bethe ansatz of \cite{Arutyunov:2004vx}.

To treat the slightly more complicated scenario of overlapping
mode indices (which can occur for three or more impurities),
we introduce the normalized eigenvectors
\be
\frac{1}{\sqrt{N_q!}}\left(a_q^\dag\right)^{N_q} a_{n_1}^\dag a_{n_2}^\dag
        \ldots a_{n_{(N_B-N_q)}}^\dag \ket{J}\ ,
\ee
which contain a single subset of $N_q$ bosonic oscillators $a_q^\dag$
which all share the same mode index $q$.  The remaining indices
$n_i \in \{n_1,n_2,\ldots,n_{N_B-N_q}\}$ are all separate from $q$
and unequal from each other,
such that the level-matching condition in eqn.~(\ref{LM}) now reads
\be
N_q\,q + \sum_{j=1}^{N_B-N_q}n_j = 0\ .
\ee
For this case we compute a matrix element analogous to that in eqn.~(\ref{delME1}):
\be
&&\kern-20pt
    \frac{1}{N_q!}\bra{J}
    \left(a_q \right)^{N_q} a_{n_1} a_{n_2}
        \ldots a_{n_{(N_B-{N_q})}}
    ( a_{-n}^\dag a_{-l}^\dag a_m a_p )
    \left(a_q^\dag\right)^{N_q} a_{n_1}^\dag a_{n_2}^\dag
        \ldots a_{n_{(N_B-{N_q})}}^\dag \ket{J}
\nn\\
&&\kern+5pt
    ={N_q}({N_q}-1)\delta_{p-q}\,\delta_{m-q}\,\delta_{n+q}\,\delta_{l+q}
    + L \sum_{j=1}^{N_B-{N_q}}
    \Bigl(
    \delta_{p-q}\,\delta_{n+q}\,\delta_{m-n_j}\,\delta_{l+n_j}
    +\delta_{m-q}\,\delta_{n+q}\,\delta_{p-n_j}\,\delta_{l+n_j}
\nn\\
&&\kern+10pt
    +\delta_{p-q}\,\delta_{l+q}\,\delta_{m-n_j}\,\delta_{n+n_j}
    +\delta_{m-q}\,\delta_{l+q}\,\delta_{p-n_j}\,\delta_{n+n_j}
    \Bigr)
    +\frac{1}{2}\sum_{j,k=1\atop j\neq k}^{N_B-{N_q}}
    \Bigl(
    \delta_{n_j+n}\,\delta_{n_k+l}\,\delta_{n_j-m}\,\delta_{n_k-p}
\nn\\
&&\kern+10pt
    +\delta_{n_j+n}\,\delta_{n_k+l}\,\delta_{n_k-m}\,\delta_{n_j-p}
    +\delta_{n_j+l}\,\delta_{n_k+n}\,\delta_{n_j-m}\,\delta_{n_k-p}
    +\delta_{n_j+l}\,\delta_{n_k+n}\,\delta_{n_k-m}\,\delta_{n_j-p}
    \Bigr)\ .
\label{delME2}
\ee
Using this result, we arrive at the $\su(2)$
energy shift for string states with $N_B$ total excitations containing
an ${N_q}$-component subset of oscillators that share the same mode index $q$:
\be
\delta E_{S^5}(\{n_i\},q,{N_q},N_B,J) & = &
    -\frac{{N_q}({N_q}-1)q^2 }{2 J \omega_q^2}
\nn\\
&&\kern-50pt
    -\sum_{j=1}^{N_B-{N_q}}
    \frac{{N_q}}{J\omega_q\omega_{n_j}}
    \left[q^2 + n_j^2(1+q^2\lambda')+q\,n_j
        \left(1-\omega_q\omega_{n_j}\lambda'\right)\right]
\nn\\
&&\kern-50pt
    -\sum_{j,k=1\atop j\neq k}^{N_B-{N_q}}
    \frac{1}{2J\,\omega_j\omega_k}\left[
    n_k^2 + n_j^2\left(1+n_k^2\lambda'\right)
    + n_j n_k\left(1-\omega_j\omega_k\lambda'\right)\right]\ .
\label{su2GENnn}
\ee

This formula can be compared with the three-impurity $\su(2)$ energy
shift with two equal mode indices ($N_q=2$) obtained in \cite{Callan:2004ev}.
For this particular case we can set the isolated mode number to $-2q$
using the level-matching condition to simplify the result:
\be
\delta E_{S^5}(q,J)  =
    -\frac{q^2}{J\omega_q^2 \omega_{2\,q} }
    \left[
    \omega_{2\,q}\left( 5+4\,q^2\lambda'\right)
    +\omega_q\left( 6+8\,q^2\lambda'\right)
    \right]\ .
\label{3impsu2nn}
\ee
It is easy to show that eqn.~(\ref{su2GENnn}) exactly reproduces
this energy shift when restricted to $N_B=3$ with a subset of two mode numbers
equal to $q$.

We now generalize the analysis completely by using eigenstates with
$M$ mode-index subsets, where all mode indices are equal within these subsets:
\be
\frac{\left( a_{q_1}^\dag\right)^{N_{q_1}}}{\sqrt{N_{q_1}!}}
\frac{\left( a_{q_2}^\dag\right)^{N_{q_2}}}{\sqrt{N_{q_2}!}}
\cdots
\frac{\left( a_{q_M}^\dag\right)^{N_{q_M}}}{\sqrt{N_{q_M}!}}\ket{J}\ .
\nn
\ee
The $j^{th}$ subset contains $N_{q_j}$ oscillators with equal mode index $q_j$,
and the total impurity number is again $N_B$, such that
\be
\sum_{i=1}^M N_{q_i} = N_B \qquad
\sum_{i=1}^M N_{q_i} q_i = 0\ .
\ee
The matrix element of $a_{-n}^\dag\, a_{-l}^\dag\, a_m\, a_p$
between the above states,
analogous to eqns.~(\ref{delME1},\ref{delME2}), is
\be
&&\kern-30pt
\bra{J}
\frac{\left( a_{q_1} \right)^{N_{q_1}}}{\sqrt{N_{q_1}!}}
\cdots
\frac{\left( a_{q_M} \right)^{N_{q_M}}}{\sqrt{N_{q_M}!}}
\left( a_{-n}^\dag\, a_{-l}^\dag\, a_m\, a_p \right)
 \frac{\left( a_{q_1}^\dag\right)^{N_{q_1}}}{\sqrt{N_{q_1}!}}
\cdots
\frac{\left( a_{q_M}^\dag\right)^{N_{q_M}}}{\sqrt{N_{q_M}!}}
\ket{J}
\nn\\
&&\kern+00pt
    = \sum_{j=1}^M N_{q_j}(N_{q_j}-1)\,\delta_{n+n_j}\,\delta_{l+n_j}\,
    \delta_{m-n_j}\,\delta_{p-n_j}
    +\frac{1}{2}\sum_{j,k=1\atop j\neq k}^M N_{q_j} N_{q_k}
    \Bigl(
    \delta_{n+n_k}\,\delta_{l+n_j}\,\delta_{m-n_k}\,\delta_{p-n_j}
\nn\\
&&\kern-10pt
    +\delta_{n+n_j}\,\delta_{l+n_k}\,\delta_{m-n_k}\,\delta_{p-n_j}
    +\delta_{n+n_k}\,\delta_{l+n_j}\,\delta_{m-n_j}\,\delta_{p-n_k}
    +\delta_{n+n_j}\,\delta_{l+n_k}\,\delta_{m-n_j}\,\delta_{p-n_k}
    \Bigr)\ .
\ee
We thereby obtain the completely general $\su(2)$ energy shift
for $N_B$-impurity string states containing $M$ equal-mode-index subsets of oscillators:
\be
\delta E_{S^5}(\{q_i\},\{N_{q_i}\},M,J) & = & -\frac{1}{2J}\biggl\{
    \sum_{j=1}^M N_{q_j}(N_{q_j}-1)
    \left(1-\frac{1}{\omega_{q_j}^2\lambda'}\right)
\nn\\
&&\kern-18pt
    +\sum_{j,k=1\atop j\neq k}^M \frac{N_{q_j}N_{q_k}}{\omega_{q_j}\omega_{q_k}}
    \left[q_k^2 + q_j^2\omega_{q_k}^2\lambda'
    +q_j q_k(1-\omega_{q_j}\omega_{q_k}\lambda')\right]\biggr\}\ .
\label{SU2FULL}
\ee
This master formula can be used to determine the $\su(2)$ string energy spectrum to
$O(J^{-1})$ for all possible physical string states in this sector.

By taking $M=2$ and setting $N_{n_1} = N_{n_2} = 1$ (using the unequal mode indices $\{n_1,n_2\}$),
we recover from this equation the exact two-impurity result recorded in eqn.~(\ref{2impsu2}) above,
with $n_2=-n_1$.
For $M=3$ and $N_{n_1} = N_{n_2} = N_{n_3} = 1$, we get
the complete three-impurity unequal-mode-number $(n_1\neq n_2\neq n_3)$
formula found in eqn.~(\ref{3impsu2}).
Finally, the three-impurity eigenvalue with two equal mode indices $(q_1=q_2,\ q_3=-2q_1)$
given in eqn.~(\ref{3impsu2nn}) can also be extracted from eqn.~(\ref{SU2FULL})
by setting $M=2$, $N_{q_1} = 2$ and $N_{q_2}=1$.

We also note that eqn.~(\ref{SU2FULL}) agrees perfectly with the corresponding
near-pp-wave formula derived from the $\su(2)$ string Bethe ansatz of \cite{Arutyunov:2004vx}
for completely general mode-number assignment.
This successful match stands as very strong evidence that their ansatz is correct,
at least to $O(J^{-1})$.

\subsection{The $SO(4)_{AdS}$ ($\Sl(2)$) sector}
Following the derivation of eqn.~(\ref{SU2FULL}) for the energy eigenvalues
of arbitrary string states in the symmetric-traceless $SO(4)_{S^5}$ sector,
it is straightforward to find the analogous expression for symmetric-traceless
string states excited in the $SO(4)_{AdS}$ subspace, dual to operators in the
$\Sl(2)$ sector of the corresponding gauge theory.  We can define, for example,
\be
a_n = \frac{1}{\sqrt{2}}\left( a^1_n + i a^2_n \right) \qquad
\bar a_n = \frac{1}{\sqrt{2}}\left( a^1_n - i a^2_n \right)\ ,
\label{oscdef2}
\ee
and carry out the above calculations by computing general matrix elements
of $ a_{-n}^\dag a_{-l}^\dag a_m a_p $ defined in terms of these oscillators.
(Here we can project onto any $(n,m)$-plane in the $AdS_5$ subspace, as long
as $n\neq m$.)
General string energy eigenvalues in the $SO(4)_{AdS}$ symmetric-traceless
irrep are thus found to be
\be
\delta E_{AdS}(\{q_i\},\{N_{q_i}\},M,J) & = & \frac{1}{2J}\biggl\{
    \sum_{j=1}^M N_{q_j}(N_{q_j}-1)
    \left(1-\frac{1}{\omega_{q_j}^2\lambda'}\right)
\nn\\
&&\kern+10pt
    +\sum_{j,k=1\atop j\neq k}^M \frac{N_{q_j}N_{q_k}}{\omega_{q_j}\omega_{q_k}}
    q_j q_k \left[1-q_j q_k\lambda'+\omega_{q_j}\omega_{q_k}\lambda'  \right]\biggr\}\ .
\label{SL2FULL}
\ee
For later reference we record the limit of this equation for states with completely unequal mode
indices ($\{N_{n_i}\}=1,\ M=N_B$):
\be
\delta E_{AdS}(\{n_i\},N_B,J) &=& \frac{1}{2J}\sum_{j,k=1\atop j\neq k}^{N_B}
    \frac{n_j n_k}{\omega_{n_j}\omega_{n_k}}\left[1-n_jn_k\lambda'
    +\omega_{n_j}\omega_{n_k}\lambda'\right]\ .
\label{sl2GEN}
\ee

When $M=2$ and $N_{n_1} = N_{n_2} = 1$ in eqn.~(\ref{sl2GEN}), we find the
two-impurity eigenvalue (with $n_2=-n_1$)
\be
\delta E_{AdS}(n_1,J) = -\frac{2\,n_1^2 \lambda'}{J}\ ,
\ee
which agrees with the two-impurity result reported in \cite{Callan:2003xr,Callan:2004uv}
(the $\su(2)$ and $\Sl(2)$ eigenvalues are degenerate in the two-impurity
regime).  For the three-impurity eigenvalue with three unequal mode indices
we set $M=3$ and $N_{n_1} = N_{n_2} = N_{n_3} = 1$ to obtain
\be
\delta E_{AdS} (n_1,n_2,n_3,J) & =  &
    \frac{1}{J\omega_{n_1}\omega_{n_2}\omega_{n_3}}\biggl\{
    n_1n_3(1-n_1n_3\lambda')\,\omega_{n_2}
    +n_1n_2(1-n_1n_2\lambda')\,\omega_{n_3}
\nn\\
&&\kern-40pt
    +\,n_2n_3(1-n_2n_3\lambda')\,\omega_{n_1}
    +\left[
    n_1n_2+n_3(n_1+n_2)\right]\lambda'
    \omega_{n_1}\omega_{n_2}\omega_{n_3}
    \biggr\}\ ,
\label{3impsl2}
\ee
which precisely reproduces the corresponding $\Sl(2)$ result reported in \cite{Callan:2004ev}.
Finally, by setting $M=2$, $N_{q_1} = 2$, $N_{q_2}=1$ and $q_1=q_2=q,\ q_3=-2q$,
eqn.~(\ref{SL2FULL}) provides the following three-impurity eigenvalue with two equal
mode indices:
\be
\delta E_{AdS}(q,J)  =
    -\frac{q^2}{J\omega_q^2 \omega_{2\,q} }
    \left[
    \omega_{2\,q}\left( 3+4\,q^2\lambda'\right)
    +\omega_q\left( 4+8\,q^2\lambda'\right)
    \right]\ .
\label{3impsl2nn}
\ee
This again matches the three-impurity formula found in \cite{Callan:2004ev}.

\subsection{The $\su(1|1)$ sector}
Based on the above results in the bosonic $SO(4)_{AdS}$ and
$SO(4)_{S^5}$ symmetric-traceless sectors, we can easily formulate
a conjecture for the $N$-impurity eigenvalue of symmetrized
pure-fermion states in either the $({\bf 2,1;2,1})$ or $({\bf
1,2;1,2})$ of $SO(4)\times SO(4)$, labelled by the $\su(1|1)$
subalgebra. We first note that, since these states are composed of
fermionic oscillators which are symmetrized in their spinor
indices, no states in this sector can carry subsets of overlapping
mode numbers (since they would automatically vanish). Furthermore,
when restricting to states with completely unequal mode indices,
we can see that the $N$-impurity eigenvalues obtained for the
$\su(2)$ and $\Sl(2)$ sectors (eqns.~(\ref{su2GEN}) and
(\ref{sl2GEN})) are obvious generalizations of the corresponding
three-impurity formulas (eqns.~(\ref{3impsu2}) and
(\ref{3impsl2}), respectively).  Namely, if the three-impurity
eigenvalues take the generic form \be \delta E(n_1,n_2,n_3,J) =
\sum_{j,k=1\atop j\neq k}^3 F(n_j,n_k)\ , \ee the $N$-impurity
generalization is simply \be \delta E(\{n_i\},N,J) =
\sum_{j,k=1\atop j\neq k}^{N} F(n_j,n_k)\ . \ee By carrying this
over to the $\su(1|1)$ sector, we find the $N$-impurity eigenvalue
of $H_{\rm FF}$ between symmetrized $({\bf 2,1;2,1})$ or $({\bf
1,2;1,2})$ fermions (the eigenvalues of both are necessarily
degenerate): \be \delta E_{\su(1|1)}(\{n_i\},N_F,J) =
-\frac{1}{4J}\sum_{j,k=1 \atop j\neq k}^{N_F}
    \frac{1}{\omega_{n_j}\omega_{n_k}}\left[n_j^2+n_k^2+2n_j^2 n_k^2 \lambda'
        -2\,n_j n_k \omega_{n_j}\omega_{n_k}\lambda' \right]\ .
\label{SU11FULL}
\ee

For $N_F=2$, this formula matches the two-impurity result in
\cite{Callan:2003xr,Callan:2004uv}:
\be
\delta E_{\su(1|1)}(n_1,J) = -\frac{2\,n_1^2 \lambda'}{J}\ ,
\ee
with $n_2=-n_1$ (this eigenvalue overlaps with the corresponding two-impurity
$\su(2)$ and $\Sl(2)$ values).  When $N_F=3$ we of course recover the
three-impurity eigenvalue reported in \cite{Callan:2004ev}:
\be
\delta E_{\su(1|1)}(n_1,n_2,n_3,J)  &=&
    -\frac{1}{4\,J\omega_{n_1}\omega_{n_2}\omega_{n_3}}\biggl\{
    -4\,\bigl(n_2n_3+n_1(n_2+n_3)\bigr)\lambda'\omega_{n_1}\omega_{n_2}\omega_{n_3}
\nn\\
&&\kern-110pt
    +\biggl[
    \omega_{n_1}
    \left(
    2\, n_3^2
    +4\,n_2^2n_3^2\lambda'
    +2\, n_2^2
    \right)
    +\bigl(n_3\to n_2,\ n_2\to n_1,\ n_1\to n_3\bigr)
    +\bigl( n_1\rightleftharpoons n_2 \bigr)
    \biggr]
    \biggr\}\ .
\nn\\
&&
\label{exactfermi}
\ee
It would be straightforward to check eqn.~(\ref{SU11FULL}) against
an explicit four-impurity calculation in the string theory, for example.
Better yet, one might carry out the direct $N$-impurity calculation in the
$H_{\rm FF}$ sector analogous to the above calculations for $H_{\rm BB}$.  The
latter would certainly be more technically
complicated than in the bosonic sectors, and for the moment we leave eqn.~(\ref{SU11FULL})
as it stands, withholding direct verification for a future study.

\section{Spectral decomposition }
At one- and two-loop order in $\lambda'$ we can infer from basic arguments
the spectral decomposition of the extended $N$-impurity superconformal
multiplet of $O(J^{-1})$ energy corrections to the pp-wave limit.
For simplicity we will restrict the discussion to eigensystems with completely unequal
mode numbers, though the generalization to more complicated cases is straightforward.
To begin we will review the two- and three-impurity supermultiplet structures
studied in \cite{Callan:2003xr,Callan:2004uv,Callan:2004ev}.

We denote the one- and two-loop energy eigenvalue shifts as
$\Lambda^{(1)}$ and $\Lambda^{(2)}$, according to the generic formula
\be
E(\{n_j\},N,J) &=& N + \frac{\lambda'}{2}\sum_{j=1}^N
    n_j^2\left(1 + \frac{\Lambda^{(1)}}{J} + O(J^{-2})\right)
\nn\\
&&  - \frac{{\lambda'}^2}{4}\sum_{j=1}^N
    n_j^4\left(\frac{1}{2} + \frac{\Lambda^{(2)}}{J} + O(J^{-2})\right)
    + O({\lambda'}^3)\ .
\label{lambda}
\ee
The fact that these energy shifts can be expressed as coefficients
of $\sum n_j^2$ and $\sum n_j^4$ is not obvious.
In the two- and three-impurity cases this was shown to be true
by direct diagonalization of the Hamiltonian.  By expanding
eqns.~(\ref{SU2FULL},\ref{SL2FULL},\ref{SU11FULL}) in small $\lambda'$,
it can also be seen that the more general $N$-impurity $\su(2)$, $\Sl(2)$
and $\su(1|1)$ eigenvalues
adhere to this structure to two-loop order.
We will argue that the remaining energy shifts (those in non-protected subsectors)
can be obtained from the protected sectors
through half-integer shifts of the $S^5$ angular momentum $J$:  it will therefore
be seen that all energies considered here will appear in the form given in
eqn.~(\ref{lambda}).

As described in \cite{Callan:2004ev}, the conformal invariance of
the full $\alg{psu}(2,2|4)$ symmetry algebra of the theory
guarantees that the energy eigenvalues (and hence $\Lambda^{(1)}$
and $\Lambda^{(2)}$) will be organized into conformal
(sub)multiplets built on conformal primary (or highest weight)
states. Within a given submultiplet we refer to states with lowest
energy as super-primary states, and the other conformal primaries
within the submultiplet are obtained by acting on super-primaries
with any of the eight supercharges, labelled by ${\cal Q}_\alpha$,
that increment $\Lambda^{(1)}$ or $\Lambda^{(2)}$ by a fixed
amount but leave the impurity number unchanged. In the gauge
theory these charges are understood to shift both the operator
dimension and $R$-charge such that $\Delta = D-R$ remains fixed
within the submultiplet. Acting with $L_{\rm sub}$ factors of
these supercharges on a super-primary generates nine levels within
each submultiplet labelled by $L_{\rm sub} = 0,\ldots,8$. If the
lowest energy level ($L_{\rm sub}=0$) in the submultiplet is
occupied by $p$ degenerate super-primaries, the $L_{\rm sub}^{th}$
level will therefore contain $p\,C_{L_{\rm sub}}^8$ degenerate
states, where $C_n^m$ is the binomial coefficient. Furthermore, if
the super-primary in a given submultiplet is a spacetime boson,
the $L_{\rm sub}={\rm even}$ levels of the submultiplet will all
be bosonic, and the $L_{\rm sub}={\rm odd}$ levels will be
fermionic.  The opposite is true if the bottom state is fermionic.

As an example, consider the one-loop, two-impurity supermultiplet
structure studied in \cite{Callan:2003xr,Callan:2004uv}. The
spectrum in this case contains only a single multiplet built on a
scalar super-primary (labelled by $1_B$, where the subscript
denotes a bosonic level) with $O(1/J)$ one-loop energy shift
$\Lambda^{(1)} = -6$. The $L_{\rm sub}=1$ level therefore has
eight degenerate states $(8_F)$ with $\Lambda^{(1)}=-5$, the
$L_{\rm sub}=2$ level contains $28_B$ states with
$\Lambda^{(1)}=-4$ and so on.  We record the two-impurity
supermultiplet structure in table~\ref{2mult1} for comparison with
higher-impurity spectra.
\begin{table}[ht!]
\begin{eqnarray}
\begin{array}{|c|ccccccccc|}
\hline
L_{\rm sub} & 0 & 1 & 2 & 3 & 4 & 5 & 6 & 7 & 8     \\  \hline
        & 1_B   & 8_F   & 28_B  & 56_F  & \fbox{$70_B$} & 56_F  & 28_B  & 8_F  & 1_B    \\ \hline
\Lambda^{(1)}(L_{\rm sub})& -6    & -5  & -4    & -3    & -2    & -1    & 0 & 1 & 2  \\     \hline
\Lambda^{(2)}(L_{\rm sub})& -4    & -3  & -2    & -1    & 0 & 1 & 2 & 3 & 4  \\     \hline
\end{array} \nonumber
\end{eqnarray}
\caption{Submultiplet breakup of the 256-dimensional two-impurity spectrum}
\label{2mult1}
\end{table}
The one-loop energies of the three protected $\Sl(2)$, $\su(2)$ and $\su(1|1)$
subsectors studied here
are degenerate in the two-impurity regime and lie in the boxed $70_B$
``centroid'' level in table~\ref{2mult1}.  We also record in table~\ref{2mult1}
the two-loop energy shifts $\Lambda^{(2)}$, which are offset from the
one-loop values by two: $\Lambda^{(2)} = \Lambda^{(1)}+2$.

In the gauge theory there are sixteen operators which increment
the impurity number by one and shift the $R$-charge by certain
amounts \cite{Beisert:2002tn}. Four of these act on
single-trace operators by rotating the $SO(6)$ scalars $Z$
(carrying one unit of $R$-charge) into $\phi$ (which carry zero
$R$-charge):  they increase the operator impurity number by one
and decrease the $R$-charge by one ($N\to N+1,\ R\to R-1$).  Four
operators rotate $Z$ into ${\cal D}Z$, increasing $N$ by one and
leaving the $R$-charge fixed.  The remaining eight operators
are fermionic and send $N\to N+1,\ R\to R+1/2$. If
one uses these operators to generate $N$-impurity super-primaries
from those in the $(N-1)$-impurity spectrum, an immediate
implication is that, within a given $N$-impurity spectrum of
anomalous dimensions, all of the eigenvalues in the gauge theory
will be related to each other by half-integer shifts in the
$R$-charge.  Certain energy levels will therefore be common to all
of the submultiplets in the spectrum built on super-primary
operators, and this special degeneracy can be used to deduce the
overall structure of the extended supermultiplet.  This
degeneracy, however, only persists in the string theory to
two-loop order in $\lambda'$, and it is for this reason that we
are forced to limit the general superstring spectral decomposition
to two-loop order in the expansion.  (It will be shown below,
however, that a certain subset of submultiplets in the string
theory can always be determined to \emph{all} orders in
$\lambda'$.)

Sending $J\to J+A$ on the string side (dual to an $R$-charge shift in the gauge theory)
shifts $\Lambda^{(1)}$ and
$\Lambda^{(2)}$ by $-2A$:  starting from the two-impurity
super-primary $(1_B)$ with energy $\Lambda^{(1)}=-6$, the string
versions of the sixteen impurity-increasing operators can be
understood to generate four (degenerate) bosonic three-impurity
super-primaries with $\Lambda^{(1)}=-8$, eight fermionic
three-impurity super-primaries with $\Lambda^{(1)}=-7$ and four
bosonic three-impurity super-primaries with $\Lambda^{(1)}=-6$. By
acting with the eight charges ${\cal Q}_\alpha$ we then generate
submultiplets based on each of these super-primaries whose levels
are populated by $p\,C_{L_{\rm sub}}^8$ degenerate states, where
$p$ here is either four (for the two four-dimensional bosonic
super-primary levels) or eight (for the eight-dimensional
fermionic super-primary level). The submultiplets themselves can
be labelled by a separate index $K$, in this case running over
$K=0,\ldots,2$.

The complete three-impurity multiplet structure is
recorded in table~\ref{3mult1}.  Here there are a total of 11 levels in the extended
supermultiplet, and we label these with the index $L$ such that $L=L_{\rm sub}+K$.
In table~\ref{3mult1} the closed $\su(2)$ sector lies in
the boxed $280_B$ level in the $K=0$ submultiplet with $\Lambda^{(1)}=-4$, the $\Sl(2)$
eigenvalue ($\Lambda^{(1)}=-2$) is in the boxed $280_B$ level of the $K=2$ submultiplet
and the $\su(1|1)$ eigenvalue ($\Lambda^{(1)}=-3$) is in the $560_F$ level of the $K=1$
submultiplet.  For any impurity number these protected eigenvalues will always
lie at the $L_{\rm sub}=4$ level within their respective submultiplets.
We also note that, in the $K$ direction, the $\su(2)$ and $\Sl(2)$ eigenvalues
will correspond to eigenstates composed purely of $S^5$ or $AdS_5$
bosonic excitations, and will therefore fall
into the ``bottom'' and ``top'' submultiplets, respectively
(the $K=0$ and $K=2$ levels in the three-impurity
case).  Similarly, the $\su(1|1)$ eigenvalue will correspond to eigenstates composed
of either $({\bf 2,1;2,1})$ or $({\bf 1,2;1,2})$ excitations, and always lie in the
``centroid'' submultiplet in the $K$ direction (the $K=1$ level for three impurities).
The energies shared by each of the submultiplets can be collected into degenerate levels of
the complete supermultiplet.  This total level degeneracy $D(L)$ is recorded in
the bottom row of table~\ref{3mult1}.
\begin{table}[ht!]
\begin{eqnarray}
\begin{array}{|c|ccccccccccc|}
\hline
K \backslash L & 0  & 1 & 2 & 3 & 4 & 5 & 6 & 7 & 8 & 9 & 10    \\  \hline\hline
0       & 4     & 32    & 112   & 224   & \fbox{$280$ }& 224    & 112   & 32    & 4 &   &   \\ \hline
1       &   & 8 & 64    & 224   & 448   & \fbox{$560$}& 448  & 224  & 64    & 8 &   \\ \hline
2       &   &   & 4     & 32    & 112   & 224   & \fbox{$280$}  & 224   & 112   & 32    & 4 \\ \hline\hline
\Lambda^{(1)}(L) & -8   & -7    & -6    & -5    & -4    & -3    & -2    & -1    & 0 & 1 & 2  \\ \hline
\Lambda^{(2)}(L) & -6   & -5    & -4   & -3 & -2    & -1    & 0 & 1 & 2 & 3 & 4  \\ \hline\hline
D(L) & 4_B  & 40_F  & 180_B & 480_F & 840_B & 1008_F& 840_B & 480_F & 180_B & 40_F  & 4_B \\
\hline
\end{array} \nonumber
\end{eqnarray}
\caption{Submultiplet breakup of the 4,096-dimensional three-impurity spectrum}
\label{3mult1}
\end{table}

It is easy to generalize this supermultiplet structure to
arbitrary impurity number based on how the complete three-impurity
spectrum is generated from the two-impurity supermultiplet above.
For $N$ impurities, the complete supermultiplet will have a total
of $16^N$ states and $5+2N$ levels: the supermultiplet level index
$L$ therefore runs over $L=0,\ldots,(4+2N)$.  The entire
supermultiplet breaks into $2N-3$ submultiplets, each of which
have nine sub-levels labelled by $L_{\rm sub}=0,\ldots,8$. The
submultiplets themselves are labelled by the index $K$, which runs
over $K=0,\ldots,(2N-4)$. The one-loop energy shifts
within the $K^{th}$ submultiplet at level $L_{\rm sub}$ are thus given by
\be
\Lambda^{(1)}_{\rm sub}(K,L_{\rm sub},N) = K+L_{\rm sub}-2(N+1)\ .
\label{lambdasub}
\ee
Equivalently, the $L^{th}$ level of the entire supermultiplet has energy shift
\be
\Lambda^{(1)}(L,N)=L-2(N+1)\ .
\label{lambdaL}
\ee
The number of degenerate states at level $L_{\rm sub}$
within the $K^{th}$ submultiplet is
\be
D_{\rm sub}(K,L_{\rm sub},N) = 4^{N-2}C^{2N-4}_{K}C_{L_{\rm sub}}^8\ ,
\ee
so that the total dimension of the $K^{th}$ submultiplet is $256\times 4^{N-2}C^{2N-4}_{K}$.
By summing the submultiplet degeneracies over a given supermultiplet level $L$,
the total number of degenerate states at level $L$ in the supermultiplet
is given (in terms of Euler's $\Gamma$ function) by
\be
D(L,N) = \frac{4^{N-2}\Gamma(2N+5)}{\Gamma(2N+5-L)\Gamma(1+L)}\ .
\ee
The level is bosonic when $L$ is even and fermionic when
$L$ is odd.  As a verification of this formula, we can check that the
total number of states in the $N$-impurity supermultiplet is indeed
\be
\sum_{L=0}^{4+2N}\frac{4^{N-2}\Gamma(2N+5)}{\Gamma(2N+5-L)\Gamma(1+L)} = 16^N\ .
\ee

As noted above, the one-loop $N$-impurity $\su(2)$ energy
corresponds to eigenstates that are composed purely of
symmetric-traceless $({\bf 1,1;2,2})$ excitations:
since each of these excitations increment the angular momentum $J$ by one,
the energy eigenvalue must
therefore lie within a submultiplet built on super-primary
states that exhibit the lowest possible energy in the extended supermultiplet.
In other words, the $\su(2)$ eigenvalue always lies
at level $L_{\rm sub}=4$ of the $K=0$ submultiplet and,
using the general formula in eqn.~(\ref{lambdasub}),
we see that it exhibits the one-loop energy shift
\be
\Lambda^{(1)}_{S^5}(N) = \Lambda^{(1)}_{\rm sub}(K=0,L_{\rm sub}=4,N) = -2(N-1)\ .
\ee
As a cross-check on this result, we note that
this agrees with the one-loop limit of the general $\su(2)$ eigenvalue formula (with unequal
mode indices) in eqn.~(\ref{su2GEN}) above (with $N_B=N$):
\be
\delta E_{S^5}(\{n_i\},N,J) = -\frac{1}{2J}
    \sum_{j,k=1\atop j\neq k}^{N} (n_j^2+n_k^2)\,\lambda' + O({\lambda'}^2)
    = -\frac{1}{J}\sum_{j=1}^{N} (N-1)\,n_j^2\, \lambda' + O({\lambda'}^2)\ .
\ee
(Note the prefactor of $1/2$ in the definition of $\Lambda^{(1)}$ in
eqn.~(\ref{lambda}).)  At this point we also see that $\Lambda^{(1)}$ indeed
appears as a coefficient of $\sum n_j^2$, as given in eqn.~(\ref{lambda}).

The $N$-impurity $\Sl(2)$ eigenvalue, composed entirely of
$({\bf 2,2;1,1})$ excitations, must lie in the ``top'' $K=2N-4$ submultiplet
at $L_{\rm sub}=4$.  This gives the one-loop energy shift
\be
\Lambda^{(1)}_{AdS}(N)
    = -2 \ .
\label{sl2oneloop}
\ee
To check this we use the general $\Sl(2)$ formula for completely unequal mode indices
in eqn.~(\ref{sl2GEN}), and again expand to one-loop order in $\lambda'$:
\be
\delta E_{AdS}(\{n_i\},N,J) &=& =\frac{1}{J}\sum_{j,k=1\atop j\neq k}^{N} n_j\,n_k\,\lambda'
    + O({\lambda'}^2)\ .
\label{sl2exp}
\ee
With the level-matching condition $\sum_{j=1}^{N} n_{j} = 0$ this becomes
\be
\delta E_{AdS}(\{n_i\},N,J)=-\frac{1}{J}\sum_{j=1}^{N} n_j^2 \lambda'+ O({\lambda'}^2) \ ,
\ee
which agrees perfectly with the prediction in eqn.~(\ref{sl2oneloop})
(and again confirms that $\Lambda^{(1)}$ here is a coefficient of $\sum n_j^2$).

Finally, the $\su(1|1)$ one-loop eigenvalue,
composed of either $({\bf 2,1;2,1})$ or $({\bf 1,2;1,2})$ spinors,
lies in the $K=N-2$ submultiplet at $L_{\rm sub}=4$,
exhibiting the one-loop energy shift
\be
\Lambda^{(1)}_{\su(1|1)}(N)
    = -N \ .
\label{su11oneloop}
\ee
Using eqn.~(\ref{SU11FULL}) we see that
\be
\delta E_{su(1|1)}(\{n_i\},N,J)
&=& -\frac{1}{4J}\sum_{j,k=1\atop j\neq k}^{N} (n_j-n_k)^2\lambda' + O({\lambda'}^2)
\nn\\
    & = & -\frac{1}{2J}\sum_{j=1}^{N} N\,n_j^2\,\lambda'\ ,
\label{su11exp}
\ee
where we have again invoked the level-matching condition to derive the last line.
This of course agrees with eqn.~(\ref{su11oneloop}).
For reference we present in table~\ref{4mult1}
the complete 65,536-dimensional four-impurity spectrum of one- and two-loop
energies.  The $\su(2)$ eigenvalue in this case lies in the
boxed $1120_B$ level with $\Lambda^{(1)} = -6$,
the $\su(1|1)$ eigenvalue is in the $6720_B$ level with $\Lambda^{(1)} = -4$,
and the $\Sl(2)$ energy lies in the $1120_B$ level with $\Lambda^{(1)} = -2$.
\begin{sidewaystable}
\begin{tabular}{|c|ccccccccccccc|}
\hline
$K$ $\backslash$ $L$    & 0&  1&   2&   3&    4&    5&    6&    7&    8&    9&    10&   11&   12 \\\hline\hline
0       & 16& 128& 448& 896&  \fbox{$1120$ }& 896&  448&  128&  16&   &     &     &     \\ \hline
1       & &  64&  512& 1792& 3584& 4480& 3584& 1792& 512&  64&   &     &       \\ \hline
2       & &  &   96&  768&  2688& 5376& \fbox{$6720$}& 5376& 2688& 768&  96&   &    \\ \hline
3       & &  &   &   64&   512&  1792& 3584& 4480& 3584& 1792& 512&  64&     \\ \hline
4       & &  &   &   &    16&   128&  448&  896&  \fbox{$1120$}& 896&  448&  128&  16   \\ \hline\hline
$\Lambda^{(1)}(L)$& -10 & -9    & -8    & -7    & -6    & -5    & -4    & -3    & -2    & -1    & 0 &1 &2 \\ \hline
$\Lambda^{(2)}(L)$& -8  & -7    & -6     & -5   & -4    & -3    & -2    & -1    & 0 & 1 & 2 &3 &4 \\ \hline\hline
$D(L)$&$16_B$& $192_F$& $1056_B$& $3520_F$& $7920_B$& $12672_F$& $14784_B$& $12672_F$&
        $7920_B$& $3520_F$& $1056_B$& $192_F$& $16_B$ \\ \hline
\end{tabular}
\caption{Submultiplet breakup of the 65,536-dimensional four-impurity spectrum}
\label{4mult1}
\end{sidewaystable}
\clearpage

Comparing the $\Lambda^{(2)}$ and $\Lambda^{(1)}$
spectra in tables~\ref{2mult1} and \ref{3mult1} (which are determined
directly from the string Hamiltonian), we see that the spectrum of $\Lambda^{(2)}$ is
identical to $\Lambda^{(1)}$ up to an overall shift.
The two-loop analogue of the general $N$-impurity energy shift of
eqn.~(\ref{lambdasub}) is therefore
\be
\Lambda^{(2)}_{\rm sub}(K,L_{\rm sub},N) = K+L_{\rm sub}-2N\ .
\label{lambda2}
\ee
Equivalently, we have $\Lambda^{(2)}(L,N) = L-2N$ for the entire supermultiplet
shift in terms of $L$.

Similar to the one-loop case, we can test this two-loop formula using
the $N$-impurity results derived above
in the three protected sectors.  According to eqn.~(\ref{lambda2}),
the $\su(2)$ eigenvalue in the $K=0$ submultiplet at level
$L_{\rm sub}=4$ has the following two-loop energy shift:
\be
\Lambda^{(2)}_{S^5} (N)
= 4-2N\ .
\ee
Isolating the two-loop energy eigenvalue $\delta E_{S^5}^{(2)}$ from the $N$-impurity
$\su(2)$ equation
(\ref{su2GEN}), we have
\be
\delta E_{S^5}^{(2)}(\{n_i\},N,J) &=& \frac{1}{4J}\sum_{j,k=1\atop j\neq k}^N(n_j^4 + n_j^3n_k
                    +n_j n_k^3 +n_k^4){\lambda'}^2
\nn\\
    &=&-\frac{1}{4J}\sum_{j=1}^N(n_j^4)(4-2N){\lambda'}^2\ ,
\ee
which matches our prediction.
The $\Sl(2)$ eigenvalue in the $K=2N-4$ submultiplet is predicted to vanish
\be
\Lambda^{(2)}_{AdS} (N)
= 0\ ,
\ee
which agrees with the two-loop expansion term in eqn.~(\ref{sl2GEN}):
\be
\delta E_{AdS}^{(2)}(\{n_i\},J) = -\frac{1}{4J}\sum_{j,k=1\atop j\neq k}^N
                    \left[n_j n_k (n_j+n_k)^2\right]{\lambda'}^2 = 0\ .
\ee
Finally, the $\su(1|1)$ pure-fermion sector in the $K=N-2$ submultiplet at $L_{\rm sub}=4$
should have an energy shift of
\be
\Lambda^{(2)}_{\su(1|1)}(N)
= 2-N\ ,
\ee
which agrees with the $\su(1|1)$ formula given in eqn.~(\ref{SU11FULL}):
\be
\delta E_{\su(1|1)}^{(2)}(\{n_i\},N,J) &=& \frac{1}{8J}\sum_{j,k=1\atop j\neq k}^N
                    (n_j^2-n_k^2)^2{\lambda'}^2
\nn\\       &=&-\frac{1}{4J}\sum_{j=1}^N(n_j^4)(2-N){\lambda'}^2\ .
\ee

As described in \cite{Callan:2004ev},
it should also be noted that since we know the $\su(2)$, $\Sl(2)$
and $\su(1|1)$ eigenvalues to all orders in $\lambda'$, we can easily
determine complete all-loop energy formulas for the three submultiplets
to which these eigenvalues belong.
It was previously noted that the eight supercharges
(${\cal Q}_\alpha$) that act as raising operators within each submultiplet
are known in the gauge theory to shift both the dimension and $R$-charge
by $1/2$ such that $\Delta = D-R$ is kept fixed.  Because all states within
a given submultiplet share the same $\Delta$, the string energy shift at
any level $L_{\rm sub}$ can therefore be obtained from that at level $L_{\rm sub}'$
by replacing
\be
J \to J-L_{\rm sub}/2 + L_{\rm sub}'/2 \nn
\ee
in the energy eigenvalue evaluated at sub-level $L_{\rm sub}'$.
Since we are expanding to $O(J^{-1})$, however, this replacement can only affect
the eigenvalues $\delta E$ via the $O(J^0)$ BMN term in the pp-wave limit.
For the protected eigenvalues
determined above at $L_{\rm sub}=4$, we therefore find the
all-loop energy shift for the entire submultiplet by including the
appropriate $O(J^{-1})$ contribution from the BMN formula
\be
E_{\rm BMN} = \sum_{j=1}^N \sqrt{1+\frac{n_j^2\lambda}{(J+2-L_{\rm sub}/2)^2}}\ .
\ee
Explicitly, the complete level spectra of the $K=0,\ K=N-2$ and $K=2N-4$
submultiplets are given, to all orders in $\lambda'$, by
\be
\delta E(\{n_j\},L_{\rm sub},N,J) = \frac{\lambda'}{2J}\sum_{j=1}^N
     \frac{n_j^2 (L_{\rm sub}-4)}{\sqrt{1+n_j^2\lambda'}}
    + \delta E_{L_{\rm sub}=4}(\{n_j\},J)\ ,
\ee where $\delta E_{L_{\rm sub}=4}$ is the $L_{\rm sub}=4$ energy
shift in the submultiplet of interest. Since the level degeneracy
among submultiplets is generally broken beyond two-loop order, it
is difficult to obtain similar expressions for submultiplets not
containing the $\su(2)$, $\Sl(2)$ and $\su(1|1)$ protected
eigenvalues. This can possibly be addressed by relying directly on
the commutator algebra of various impurity-increasing operators in
the string theory, and we will return to this problem in a future
study.

\section{Discussion}
We have directly computed the near-pp-wave eigenvalues of
$N$-impurity bosonic string states with arbitrary mode-number
assignment lying in the protected symmetric-traceless irreps of
the $AdS_5$ ($\Sl(2)$) and $S^5$ ($\su(2)$) subspaces. Based on
the observation that the $\su(2)$ and $\Sl(2)$ eigenvalues are
simple generalizations of the three-impurity results obtained in
\cite{Callan:2004ev}, we have also presented a conjecture for the
$N$-impurity eigenvalues of symmetrized-fermion states in the
$\su(1|1)$ sector. This conjecture meets several basic
expectations and we believe that it is correct. (It would be
satisfying, however, to derive the $\su(1|1)$ eigenvalue formula
directly from the fermionic sector of the string theory.) We have
also found that the $\su(2)$ eigenvalues perfectly match, to all
orders in $\lambda'$, the corresponding eigenvalue predictions
given by the string Bethe ansatz of \cite{Arutyunov:2004vx}.
Along these lines, it would be very interesting to have long-range Bethe ans\"atze
analogous to the string \cite{Arutyunov:2004vx}
and gauge theory \cite{Serban:2004jf,Beisert:2004hm}
$\su(2)$ equations for either the protected
$\Sl(2)$ and $\su(2|3)$ sectors or for the entire $\alg{psu}(2,2|4)$
algebra of the theory.

The supermultiplet decomposition given in section 4 is based
on the breakup of the energy spectrum observed
between the two- and three-impurity regime, and is precisely what
is expected from the gauge theory based on how sixteen particular charges are
known to act on operators which are dual to the string states
of interest \cite{Callan:2004ev,Beisert:2002tn}.
Assuming that this mechanism is not specific to the
three-impurity case, we were able to generalize the decomposition
of the $N$-impurity (unequal mode index) supermultiplet to two-loop
order in $\lambda'$.  By knowing where the eigenvalues of the $\su(2)$, $\Sl(2)$ and
$\su(1|1)$ sectors are supposed to appear in this decomposition, we
were able to provide a stringent cross-check of our results,
and we have found perfect agreement.  Given the many implicit
assumptions in this procedure, however, it would be instructive
to perform a direct diagonalization of the four-impurity Hamiltonian
to test our predictions.  While such a test is likely to be computationally
intensive, the problem could be simplified to some extent by restricting
to the pure-boson $H_{\rm BB}$ sector at one loop in $\lambda'$.
We of course expect complete agreement with the results presented here.

\section*{Acknowledgements}
We would like to thank Curtis Callan and John Schwarz for support
and guidance.  IS thanks both the James A.~Cullen Memorial Fund
for support, and the organizers and participants of the 2004 PiTP
summer program for many enlightening discussions. This work was
supported in part by US Department of Energy grant
DE-FG03-92-ER40701.

\end{document}